\documentclass[twocolumn,showpacs]{revtex4}

\usepackage[dvips]{graphicx}
\usepackage{amsmath}
\usepackage{subfigure}
\usepackage{layout}
\newcommand{\bigcell}[2]{\begin{tabular}{@{}#1@{}}#2\end{tabular}}
\usepackage{array}
\usepackage{color}

\newcolumntype{M}{>{\centering\arraybackslash}m{2.5cm}}
\newcolumntype{S}{>{\centering\arraybackslash}m{1.5cm}}

\begin{document}

\title{Emission of OAM entangled photon pairs in a nonlinear ring fiber utilizing spontaneous
parametric down-conversion}

\author{D. Jav\r{u}rek}
\email{javurek@slo.upol.cz}
\address{RCPTM, Joint Laboratory of Optics of Palack\'y
University and Institute of Physics of AS CR, 17. listopadu
12, 771 46 Olomouc, Czech Republic}

\author{J. Svozil\'{i}k}
\address{RCPTM, Joint Laboratory of Optics of Palack\'y
University and Institute of Physics of AS CR, 17. listopadu
12, 771 46 Olomouc, Czech Republic}
\address{ICFO-Institut de Ci{\`{e}}ncies Fot{\`{o}}niques,
Mediterranean Technology Park, Av. Carl Friedrich Gauss 3,
08860 Castelldefels, Barcelona, Spain}

\author{J. Pe\v{r}ina Jr.}
\address{Institute of Physics, Joint Laboratory of Optics of
Palack\'y University and Institute of Physics of AS CR, 17.
listopadu 50a, 771 46 Olomouc, Czech Republic}

\begin{abstract}
We suggest the generation of photon pairs in a thermally induced
nonlinear periodically-poled silica fiber by spontaneous
parametric down-conversion. Photons are generated directly in
eigenstates of optical angular momentum. Photons in a pair can be
entangled in these states as well as in frequencies. We identify
suitable spatial and polarization modes giving an efficient
nonlinear interaction. By changing the pump field properties both
narrow- and broad-band down-converted fields can be obtained.
\end{abstract}



\maketitle

\section{Introduction}

Entangled photon fields represent a corner-stone for various
experimental implementations of quantum systems due to their
highly nonclassical behavior. Entanglement of photons is crucial
in many applications including quantum computing
\cite{Nielsen2010Quantum,Zhang2007Demonstration}, quantum
metrology \cite{Migdall1999metrology,Fraine2012dispersion} and
quantum object identification \cite{Uribe-Patarroyo2013oam}. It is
important also in the area of quantum random walks where it brings
new dimension into the problem
\cite{Shenvi2003Quantum,Svozilik2012Implementation}. Entanglement
is indispensable also for quantum communication protocols that,
among others, include quantum teleportation
\cite{Bouwmeester1997Experimental} and secured quantum key
distribution networks \cite{Ribordy2000Long,Chapuran2009Optical}.
Systems for wavelength-division-multiplexing that allow to
distribute polarization entangled photons among multiple users
\cite{Lim2008Wavelength,Svozilik2011Generation} serve as typical
examples.

In the above described applications, entangled photon pairs
generated in a nonlinear medium with non-zero $\chi^{(2)}$
susceptibility via the process of spontaneous parametric
down-conversion (SPDC) \cite{Louisell1961spdc,
Magde1967spdc,Harris1967spdc} serve as typical resources. The
entangled fields called signal and idler emerge instantly after
the annihilation of a pump photon. The nonlinear process
encompassing the pump, signal and idler photons satisfies the
conservation law of energy and phase-matching (PM) conditions
\cite{Boyd2008}. Whereas the conservation law of energy originates
in homogenity of time and is automatically fulfilled, the
phase-matching conditions for wave vectors cannot be satisfied in
a typical material with normal dispersion. However, this problem
can be overcome by utilizing birefringent materials or periodical
poling of nonlinear materials
\cite{Brida2009qpm,Svozilik2009qpm,Helmy2011Recent}. Also, the
modal PM found in waveguides allows to compensate additionally the
naturally occurring phase mismatch \cite{Helmy2007Phase}.
Alternatively, wide possibilities for achieving PM are offered by
nonlinear photonic structures
\cite{PerinaJr2006pbg,PerinaJr2011pbg}.

The problem of phase matching occurs also in the considered
nonlinear thermally poled silica fibers in which their material
dispersion is modified by their geometry (waveguiding dispersion).
Fortunately, the process of periodical poling of $ \chi^{(2)} $
susceptibility \cite{Fejer1992,PhanHuy2007photon,Zhang2008poling}
for these fibers has been mastered. It allows to achieve phase
matching of the nonlinearly interacting fields, together with the
conservation law of energy. During the poling, a SiO$ _2$ material
with no natural $\chi^{(2)}$ susceptibility (due to symmetry) is
heated up and exposed to a strong electric field originating in
the electrodes inserted in the fiber. The free ions in the fiber
are dragged by the field and form the macroscopic charge nearby
the electrodes. When the material is cooled down back to the room
temperature, the electric field is switched off. However, the ions
remain frozen at their positions and form a permanent internal
static electric field \cite{Zhang2008poling,Canagasabey2009shg}.
This field is responsible for nonlinear properties of SiO$ _2 $.
The nonlinear grating is created by a UV erasure process that
removes the nonlinearity inside domains exposed to a UV laser. The
created nonlinear grating provides an additional momentum to phase
matching conditions of the nonlinear interaction. Suitable choice
of this momentum then allows to reach quasi-phase-matching (QPM)
of the overall nonlinear process.

In the considered optical fiber, we concentrate our attention to
guided modes with a defined mode of orbital angular momentum (OAM)
\cite{Euser2008,Ramachandran2009,Bozinovic2013Terabit,Yue2012mode}.
Such modes have been theoretically studied
\cite{Bozinovic2013Terabit,Yue2012mode} and experimentally
characterized recently \cite{Bozinovic2013Terabit} using a ring or
vortex geometry (concentric rings) of optical fibers. This
geometry allows to separate modes of the LP$_{11}$ family that
differ in their effective refractive indices. This results in
their stable propagation with a low ratio of crosstalk for lengths
over 1~km. Such stable states of OAM can then be exploited to
multiplex data and rise the transfer capacity of channels.

Entangled photon pairs can alternatively be generated from
other sources based on $ \chi^{(2)} $ nonlinearity. Planar
(rectangular) periodically-poled waveguides made of PPKPT
or LiNbO$_3$
\cite{Silberhorn2009,Herrmann2013,Harder2013,Jachura2014}
represent well-developed and highly-efficient photon-pair
sources. However, their mode profiles reflect their
rectangular transverse profiles that cannot be easily and
effectively transformed into modes of fibers. On the other
hand, there exist structured photonic waveguides with
eigenmodes close to OAM modes
\cite{Euser2008,Ramachandran2009}. Unfortunately, these
fibers have transverse profiles typically few tens of
micrometers wide and so they are not suitable for thermal
poling. Thus, the investigated ring shaped fibers, though
only weakly nonlinear \cite{Zhu2012direct}, may provide an
interesting solution to the problem. Also the weakness of
nonlinear interaction may partly be compensated by the
fiber length.

OAM fields are beneficial for both the classical and quantum areas
of physics. Sufficiently strong (classical) fields are namely
useful for nano-particle manipulations \cite{Padgett2011tweezers}.
From the point of view of quantum communications that use
individual photon pairs, entangled states are crucial. As we show
below, the process of SPDC in the thermally poled fibers discussed
above allows to generate photon pairs entangled in different
degrees of freedom. These fibers then represent sources of
entangled photons that can be directly integrated into optical
fiber networks \cite{Zhu2012direct}. We note that entanglement in
OAM numbers offers additional advantage for the construction 
division multiplexing systems \cite{Bozinovic2013Terabit}. OAM
multi/demultiplexers have been recently addressed in
\cite{Guan2014Free}. Efficiency of these systems has been
characterized via the crosstalk between demultiplexed OAM modes
(the maximum value equaled -8 dB) and total losses ($\sim18$ dB).
The entangled OAM fields also allow to implement various quantum
computation protocols including the above mentioned quantum random
walks \cite{Hamilton2011Quantum,Goyal2013Implementing} and a CNOT
gate \cite{Deng2007Quantum}. Last but not least, OAM fields have
been found extraordinarily useful in the area of atomic physics
where they enable enhanced control of transitions between atomic
levels \cite{Schmiegelow2012Light}.

Photon pairs in fibers can also be generated via the
process of four-wave mixing using $ \chi^{(3)} $
susceptibility available in usual optical fibers. However,
there also occur other competing nonlinear processes based
on $ \chi^{(3)} $ susceptibility (Raman scattering). Their
presence results in larger values of single-photon noise
superimposed on photon-pair fields. Despite this, a lot of
attention has been devoted to such sources emitting photon
pairs both around 800~nm and 1550~nm
\cite{Li2005optical,Fulconis2005,Fan2005}.

The paper is organized as follows. In Sec.~II, a
theoretical model of SPDC in a ring fiber is developed
using fiber eigenmodes and propagation constants. Sec.~III
describes the decomposition into OAM modes and Sec.~IV.
brings analysis of eigenmodes of a ring fiber. Generation
of photon pairs with nonzero OAM numbers are discussed in
Sec.~V. Sec.~VI is devoted to the generation of
wide-band down-converted fields. Suitable conditions for the
generation of photon pairs entangled in OAM states are
analyzed in Sec.~VII. Also quantification of entanglement
of photons in a pair is provided in this section. Sec.~VIII
brings conclusions.

\section{Spontaneous parametric down-conversion in a ring fiber}

Nonlinear process of SPDC occurring among the pump ($ p $),
signal ($ s $) and idler ($ i $) fields can be described in
general by the following interaction Hamiltonian $
\hat{H}_{\rm int} $
\cite{Mandel1995}:
\begin{eqnarray}   
 \hat{H}_{\rm int}(t) &=& 2\varepsilon_0 \int_{S_{\perp}} rdr \,
  d\theta \int_{-L}^{0} dz\, \mathbf{\chi}^{(2)}(z):
  \mathbf{E}_{p}^{(+)}(r,\theta,z,t) \nonumber \\
 & & \times \hat{\mathbf{E}}_{s}^{(-)}(r,\theta,z,t)
  \hat{\mathbf{E}}_{i}^{(-)}(r,\theta,z,t) + {\rm
 h.c.}
\label{1}
\end{eqnarray}
Symbol : means tensor shorthand with respect to its three indices,
$ \varepsilon_0 $ denotes the vacuum permittivity and $ {\rm h.c.}
$ replaces the Hermitian conjugated term. A vector
positive-frequency electric-field amplitude of a pump beam is
denoted as $\mathbf{E}_{p}^{(+)}(r,\theta,z,t) $ whereas vector
negative-frequency electric-field operator amplitudes of the
signal and idler beams are described as $
\hat{\mathbf{E}}_{s}^{(-)}(r,\theta,z,t) $ and $
\hat{\mathbf{E}}_{i}^{(-)}(r,\theta,z,t) $, respectively.
Nonlinear susceptibility $ \mathbf{\chi}^{(2)} $ is assumed
$z$-dependent. Its spatial periodic rectangular modulation along
the $z$ axis with certain period permits quasi-phase-matching of
the nonlinear process. Hamiltonian $ \hat{H}_{\rm int} $ in
Eq.~(\ref{1}) is written in cylindrical coordinates with radial
variable $ r $, angular variable $ \theta $ and longitudinal
variable $ z $. Symbol $S_{\perp}$ denotes the transverse area of
the fiber of length $L$.

Thermal poling of SiO$_2$ material giving nonlinearity to
the fiber results in the following non-zero elements of
$\mathbf{\chi}^{(2)}$ tensor: $\chi^{(2)}_{xxx} \simeq
3\chi^{(2)}_{xyy}$ and
$\chi^{(2)}_{xyy}=\chi^{(2)}_{yyx}=\chi^{(2)}_{yxy}= 0.021$~pm/V
\cite{Boyd2008,Zhu2010measurement}. These values of $
\chi^{(2)} $ nonlinearity are approx. three orders in
magnitude lower than those characterizing LiNb0$_3$, the
most-frequently used nonlinear
material ($\chi^{(2)}\approx 30 $~pm/V). On the other hand,
sufficiently long fibers allow, at least partially, to compensate
the weak nonlinearity \cite{Canagasabey2009shg}. Also, optical
fibers profit from the transverse confinement of the guided
fields. It is important to note that the ring fibers partially
loose their radial symmetry owing to the presence of two thin
metallic wires used for thermal poling. However, the holes with
wires are usually far from the fiber core and so their influence
to radial symmetry of fiber modes results in only weak anisotropy
that can usually be omitted.

In the considered ring fiber with its rotational symmetry around
the $ z $ axis, the pump, signal and idler fields can be
decomposed into transverse eigenmodes $ {\bf
e}_\eta(r,\theta,\omega) $ with propagation constants $
\beta_\eta(\omega) $ at the appropriate frequencies $ \omega $.
Multi-index $ \eta $ contains a mode name \cite{Yeh2010waveguides}
including azimuthal ($ n $) and radial indices and polarization
index $ \phi $. In this decomposition, the strong (classical)
positive-frequency electric-field pump amplitude $ {\bf E}_p^{(+)}
$ attains the form
\begin{eqnarray}  
 {\bf E}_p^{(+)}(r,\theta,z,t) &=& \sum_{\eta_p} A_{p,\eta_p} \int d\omega_p\,
  {\cal E}_p(\omega_p) {\bf e}_{p,\eta_p}(r,\theta,\omega_p) \nonumber \\
 & & \mbox{} \times \exp\left[i\beta_{p,\eta_p}(\omega_p)z-i\omega_p
  t\right],
\label{2}
\end{eqnarray}
in which $ A_{p,\eta_p} $ gives the amplitude of mode $ \eta_p $
and $ {\cal E}_p $ stands for the pump normalized amplitude
spectrum. As the normalized eigenmodes $ {\bf
e}_\eta(r,\theta,\omega) $ form a basis, they can be used for
quantization of the signal- and idler-field photon fluxes
\cite{Vogel2001,Huttner1990}. As a consequence, the
negative-frequency electric-field signal and idler operator
amplitudes $ \hat{\bf E}_{s}^{(-)} $ and $ \hat{\bf E}_{i}^{(-)} $
can be expressed as
\begin{eqnarray}  
 \hat{\bf E}_{a}^{(-)}(r,\theta,z,t)&=& \sum_{\eta_a} \int d\omega_a
  \sqrt{\frac{\hbar \omega_a }{ 4\pi\varepsilon_{0} c \bar{n}_{a,\eta_{a}}} } \,
  \hat{a}_{a,\eta_{a}}^\dagger(\omega_a) \nonumber\\
 & & \hspace{-26mm} \times {\bf e}_{a,\eta_a}^{*}(r,\theta,\omega_a)
 \exp\left[i\beta_{a,\eta_a}(\omega_a)z-i\omega_a
  t\right], \hspace{1mm} a=s,i;
\label{3}
\end{eqnarray}
$ \hbar $ is the reduced Planck constant, $ c $ speed of light in
the vacuum and $ \bar{n}_{a,\eta_{a}} $ effective index of
refraction for mode $\eta_a$ of field $ a $ ($
\bar{n}_{a,\eta_{a}} = c\beta_{a,\eta_a}/\omega_a $). The boson
creation operators $ \hat{a}_{a,\eta_{a}}^\dagger(\omega_a) $ in
Eq.~(\ref{3}) add one photon into mode $ a $ with index $ \eta_a $
and frequency $ \omega_a $. We note that the eigenmodes are
normalized such that $ \int rdrd\theta \, |{\bf
e}_{a,\eta_a}(r,\theta,\omega_a)|^2 = 1 $.

\begin{figure}
\hspace{0.05cm}
 \subfigure[]{\includegraphics[width=3.75cm]{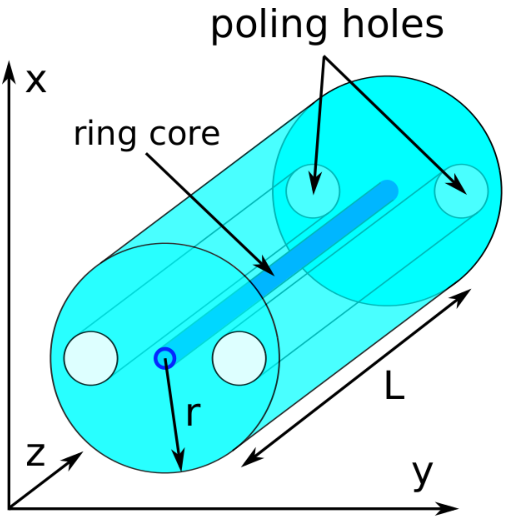}}%
 \hspace{0.1cm}
 \subfigure[]{\includegraphics[width=4.5cm]{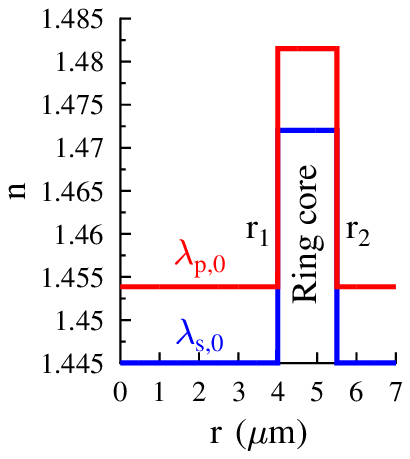}}
 \caption{[Color online:] (a) Sketch of a ring fiber with two small poling holes and
  (b) radial profiles of indices of refraction $n$ at
  the pump ($\lambda_{p}^{0}= 0.775 $ $\mu m$) and signal ($\lambda_{s}^{0}=1.55$ $\mu m$) wavelengths.}
\label{fig1}
\end{figure}

For the considered ring fiber composed of SiO$_2$ cladding and
SiO$_2$ core doped by $19.3$ mol\% of GeO$_2$ (for the scheme, see
Fig.~\ref{fig1}) \cite{Bruckner2011elements,Huynh2004}, the
normalized electric-field eigenmodes $ {\bf
e}_\eta(r,\theta,\omega) $ together with the accompanying
normalized magnetic-field eigenmodes $ {\bf
h}_\eta(r,\theta,\omega) $ can be obtained analytically.
Their longitudinal $ z $ components can be expressed in terms of
Bessel functions of the first ($ J_n $) and second ($ Y_n $) kind
and modified Bessel functions of the first ($ I_n $) and second ($
K_n $) kind as follows \cite{Snyder1983}:
\begin{eqnarray}  
 {\bf e}_{z,\eta}(r,\theta,\omega) &=& \Bigl\{ C_\eta^{(0)}(\omega)
  I_n(w_\eta^{(0)}r) {\rm rect}_{0,r_1}(r)  \nonumber \\
 & & \hspace{-20mm} \mbox{} + \left[ C_\eta^{(1)}(\omega) J_n(w_\eta^{(1)}r) +
   D_\eta^{(1)}(\omega) Y_n(w_\eta^{(1)}r) \right] {\rm rect}_{r_1,r_2}(r) \nonumber \\
 & & \hspace{-15mm} \mbox{} D_\eta^{(2)}(\omega) K_n(w_\eta^{(2)}r) {\rm rect}_{r_2,\infty}(r)
  \Bigr\} \sin(n\theta + \phi) , \nonumber \\
 {\bf h}_{z,\eta}(r,\theta,\omega) &=& \Bigl\{ A_\eta^{(0)}(\omega)
  I_n(w_\eta^{(0)}r) {\rm rect}_{0,r_1}(r)  \nonumber \\
 & & \hspace{-20mm} \mbox{} + \left[ A_\eta^{(1)}(\omega) J_n(w_\eta^{(1)}r) +
   B_\eta^{(1)}(\omega) Y_n(w_\eta^{(1)}r) \right] {\rm rect}_{r_1,r_2}(r) \nonumber \\
 & & \hspace{-15mm} \mbox{} B_\eta^{(2)}(\omega) K_n(w_\eta^{(2)}r) {\rm rect}_{r_2,\infty}(r)
  \Bigr\} \cos(n\theta + \phi) .
\label{4}
\end{eqnarray}
Function $ {\rm rect}_{a,b}(r) $ equals 1 for $ r \in <a,b> $ and
is zero otherwise. Whereas the Bessel functions describe the
oscillating solutions inside the ring core with higher index of
refraction extending from $ r=r_1 $ to $ r=r_2$, the modified
Bessel functions express the exponentially growing solutions in
the center of the fiber and the exponentially decreasing solutions
in the outer cladding. Transverse components of the wave vector $
w^{(q)} $ introduced in Eqs.~(\ref{4}) are real and they differ
according to the radial region:
\begin{eqnarray} 
 w_\eta^{(q)}(\omega) &=& \sqrt{\beta_\eta^2(\omega) -
  \frac{\omega^2}{c^2}\varepsilon_r^{(q)}(\omega)} , \hspace{5mm} q=0,2,
  \nonumber \\
 w_\eta^{(1)}(\omega) &=& \sqrt{
 \frac{\omega^2}{c^2}\varepsilon_r^{(1)}(\omega) - \beta_\eta^2(\omega)}.
\label{5}
\end{eqnarray}
Relative permittivity $ \varepsilon_r^{(1)}(\omega) $
characterizes the fiber ring core, permittivity $
\varepsilon_r^{(0)}(\omega) $ describes the fiber inner cladding
and permittivity $ \varepsilon_r^{(2)}(\omega) $ is appropriate
for the fiber outer cladding. All permittivities are considered to
be scalar quantities. Values of real coefficients $
A_\eta^{(0)}(\omega) $, $ A_\eta^{(1)}(\omega) $, $
B_\eta^{(1)}(\omega) $, $ B_\eta^{(2)}(\omega) $, $
C_\eta^{(0)}(\omega) $, $ C_\eta^{(1)}(\omega) $, $
D_\eta^{(1)}(\omega) $, and $ D_\eta^{(2)}(\omega) $ occurring in
Eqs.~(\ref{4}) are obtained from the requirement of continuity of
tangential ($ \theta $ and $ z $) components of vector electric-
[${\bf e}_\eta(r,\theta,\omega) $] and magnetic-field [${\bf
h}_\eta(r,\theta,\omega) $] amplitudes at the boundaries of the ring core.
This continuity requirement is fulfilled only for specific values
of the propagation constant $ \beta_{\eta}(\omega) $ that arise as
the solution of dispersion equation
\cite{Snyder1983,Yeh2010waveguides}.

The $ \theta $ and $ r $ components of the electric- and
magnetic-field amplitudes are obtained from their $ z $ components
in Eqs.~(\ref{4}) using the following formulas originating in the
Maxwell equations,
\begin{eqnarray}   
 {\bf e}_{r,\eta} &=& \frac{c^2}{\varepsilon_r
  \omega^2-\beta_\eta^2 c^2} \left[ \frac{i\omega\mu_0}{r}
  \frac{\partial {\bf h}_{z,\eta}}{\partial
  \theta} + i\beta_\eta \frac{\partial {\bf e}_{z,\eta}
   }{\partial r} \right], \nonumber \\
 {\bf e}_{\theta,\eta} &=& \frac{c^2}{\varepsilon_r
  \omega^2-\beta_\eta^2 c^2} \left[ -i\omega\mu_0
  \frac{\partial {\bf h}_{z,\eta}}{\partial r}
  + \frac{i\beta_\eta}{r} \frac{\partial {\bf e}_{z,\eta}
   }{\partial \theta} \right], \nonumber \\
 {\bf h}_{r,\eta} &=& \frac{c^2}{\varepsilon_r
  \omega^2-\beta_\eta^2 c^2} \left[ - \frac{i\omega\varepsilon_0\varepsilon_r}{r}
  \frac{\partial {\bf e}_{z,\eta}}{\partial
  \theta} + i\beta_\eta \frac{\partial {\bf h}_{z,\eta}
   }{\partial r} \right] , \nonumber \\
 {\bf h}_{\theta,\eta} &=& \frac{c^2}{\varepsilon_r
  \omega^2-\beta_\eta^2 c^2} \left[ i\omega\varepsilon_0\varepsilon_r
  \frac{\partial {\bf e}_{z,\eta}}{\partial
  r} + \frac{i\beta_\eta}{r} \frac{\partial {\bf h}_{z,\eta}
   }{\partial \theta} \right] .
\label{6}
\end{eqnarray}
Alternatively, the $ \theta $ and $ r $ components can be replaced
by the cartesian $ x $ and $ y $ components obtained by the simple
relations:
\begin{eqnarray}    
 {\bf e}_{x,\eta}(r,\theta,\omega) &=& \cos(\theta) {\bf e}_{r,\eta}(r,\theta,\omega)
  - \sin(\theta) {\bf e}_{\theta,\eta}(r,\theta,\omega) , \nonumber \\
 {\bf e}_{y,\eta}(r,\theta,\omega) &=& \sin(\theta) {\bf e}_{r,\eta}(r,\theta,\omega)
  + \cos(\theta) {\bf e}_{\theta,\eta}(r,\theta,\omega) .  \nonumber \\
 & &
\label{7}
\end{eqnarray}

The electric- and magnetic-field amplitudes for azimuthal index $
n \ne 0 $ in Eqs.~(\ref{4}) also depend on phase $ \phi $ that
determines the mode polarization. An eigenmode with vertical
(horizontal) polarization $ V $ ($ H $) is obtained for $ \phi = 0
$ ($ \phi = \pi/2 $). As pairs of eigemodes with $ V $ and $ H $
polarizations have the same propagation constant $ \beta_\eta $,
eigenmodes with right- ($ R $) and left-handed ($ L $) circular
polarizations can be built from these eigemodes using the
relations:
\begin{eqnarray}   
 {\bf e}_{z,\tilde{\eta} R}(r,\theta,\omega) &=& \frac{1}{\sqrt{2}}
  \left[{\bf e}_{z,\tilde{\eta} V}(r,\theta,\omega) -i
   {\bf e}_{z,\tilde{\eta} H}(r,\theta,\omega) \right] ,
   \nonumber \\
 {\bf e}_{z,\tilde{\eta} L}(r,\theta,\omega) &=& \frac{1}{\sqrt{2}}
  \left[{\bf e}_{z,\tilde{\eta} V}(r,\theta,\omega) +i
   {\bf e}_{z,\tilde{\eta} H}(r,\theta,\omega) \right], \nonumber \\
 & &
\label{8}
\end{eqnarray}
where $\tilde{\eta}$ indicates a mode excluding its polarization.
These eigemodes are close to OAM eigemodes and in general posses
nonzero OAM numbers. The electric- and magnetic-field amplitudes
for $ n = 0 $ in Eqs.~(\ref{4}) describe two orthogonal TE$ _{01}
$ and TM$ _{01} $ eigenmodes with different propagation constants
$ \beta_{\eta}(\omega) $. Polarization of TE$_{01}$ [TM$_{01}$]
mode is obtained for $\phi=0$ [$\phi=\pi/2$].

A common state $ |\psi\rangle $ of the signal and idler fields at
the output face of the fiber describing one photon pair is
determined by a first-order perturbation solution of the
Schr\"{o}dinger equation with the interaction Hamiltonian $
\hat{H}_{\rm int} $,
\begin{equation}  
 |\psi \rangle = -\frac{i}{\hbar} \int_{-\infty}^{\infty} dt
  \, \hat{H}_{\rm int}(t) |{\rm vac} \rangle.
 \label{9}
\end{equation}
State $ |{\rm vac} \rangle $ denotes the vacuum state.

Substitution of the expressions from Eqs.~(\ref{1}--\ref{3}) into
Eq.~(\ref{9}) provides the output state $ |\psi\rangle $ in the
form:
\begin{eqnarray}  
 |\psi\rangle &=& \sum_{\eta_p} \sum_{\eta_s,\eta_i}
  \int d\omega_s \int d\omega_i \, \Phi_{\eta_s\eta_i}^{\eta_p}(\omega_s,\omega_i)
   \nonumber \\
 & & \mbox{} \times \hat{a}^{\dagger}_{s,\eta_s}(\omega_s) \hat{a}^{\dagger}_{i,\eta_i}(\omega_i)
  |{\rm vac} \rangle.
 \label{10}
\end{eqnarray}
Two-photon spectral amplitudes $
\Phi_{\eta_s\eta_i}^{\eta_p}(\omega_s,\omega_i) $ introduced in
Eq.~(\ref{10}) give a probability amplitude of generating a signal
photon into mode $ \eta_s $ with frequency $ \omega_s $ together
with an idler photon into mode $ \eta_i $ with frequency $
\omega_i $ from a pump photon in mode $ \eta_p $. They are derived
as follows:
\begin{equation} 
 \Phi_{\eta_s\eta_i}^{\eta_p}(\omega_s,\omega_i) =
  - \frac{i\sqrt{\omega_s \omega_i}}{\sqrt{\bar{n}_{s,\eta_s}\bar{n}_{i,\eta_i}} c} A_{p,\eta_p}
  {\cal E}_p(\omega_s+\omega_i)
  I^{\eta_p}_{\eta_s\eta_i}(\omega_s,\omega_i) ,
\label{11}
\end{equation}
where
\begin{eqnarray}   
 I^{\eta_p}_{\eta_s\eta_i}(\omega_s,\omega_i) &=& \sqrt{2\pi}
  \int_{S_{\perp}} rdr d\theta \, \tilde{\mathbf{\chi}}^{(2)}
   [-\Delta\beta^{\eta_p}_{\eta_s\eta_i}(\omega_s,\omega_i)] \nonumber \\
 & & \hspace{-25mm} \mbox{} : {\bf e}_{p,\eta_p}(r,\theta,\omega_s+\omega_i)
  {\bf e}_{s,\eta_s}^{*}(r,\theta,\omega_s) {\bf e}_{i,\eta_i}^{*}(r,\theta,\omega_i)
\label{12}
\end{eqnarray}
and $\Delta\beta^{\eta_p}_{\eta_s\eta_i}(\omega_s,\omega_i) =
\beta_{p,\eta_p}(\omega_s+\omega_i)-\beta_{s,\eta_s}(\omega_s)-\beta_{i,\eta_i}(\omega_i)
$ characterizes phase mismatch of the nonlinear interaction.

Fourier transform $ \tilde{\mathbf{\chi}}^{(2)}(\beta) $ of spatially
modulated $ \mathbf{\chi}^{(2)}(z) $ nonlinearity used in
Eq.~(\ref{12}) is given as follows:
\begin{equation} 
 \tilde{\mathbf{\chi}}^{(2)}(\beta) = \frac{1}{\sqrt{2\pi}}
  \int_{-\infty}^{\infty} dz \mathbf{\chi}^{(2)}(z) \exp(-i\beta z) .
\label{13}
\end{equation}
It attains the following form for the considered rectangular
modulation composed of $ 2N+1 $ periods of length $ \lambda $ [$ L
= (2N+1)\Lambda $]:
\begin{eqnarray} 
 \tilde{\mathbf{\chi}}^{(2)}(\beta) &=& \mathbf{\chi}^{(2)} \, \frac{2}{
  \sqrt{2\pi}\beta} \sin(\beta\Lambda/4)
  \frac{\sin[(N+1/2)\beta\Lambda]}{\sin(\beta\Lambda/2)}
  \nonumber \\
 & & \mbox{} \times \exp(i\beta\Lambda/4) \exp(iN\beta\Lambda).
\label{14}
\end{eqnarray}

Photon-pair number density $
N_{\eta_s\eta_i}^{\eta_p}(\omega_s,\omega_i) $ belonging to an
individual nonlinear process $ (\eta_p,\eta_s,\eta_i) $ is defined
as
\begin{equation}   
 N_{\eta_s\eta_i}^{\eta_p}(\omega_s,\omega_i) = \langle \psi|
  \hat{a}_{s,\eta_s}^\dagger(\omega_s) \hat{a}_{i,\eta_i}^\dagger(\omega_i)
  \hat{a}_{s,\eta_s}(\omega_s)\hat{a}_{i,\eta_i}(\omega_i)
  |\psi\rangle .
\label{15}
\end{equation}
Using Eq.~(\ref{10}), the density $ N_{\eta_s\eta_i}^{\eta_p} $
can be expressed in a simple form:
\begin{equation}  
 N_{\eta_s\eta_i}^{\eta_p}(\omega_s,\omega_i) =
  |\Phi_{\eta_s\eta_i}^{\eta_p}(\omega_s,\omega_i)|^2.
\label{16}
\end{equation}
The corresponding signal photon-number density $
N_{s,\eta_s\eta_i}^{\eta_p}(\omega_s) $ is then derived according
to the formula
\begin{equation}  
 N_{s,\eta_s\eta_i}^{\eta_p}(\omega_s) = \int d\omega_i \,
  N_{\eta_s\eta_i}^{\eta_p}(\omega_s,\omega_i).
\label{17}
\end{equation}

Whereas the two-photon amplitudes $ \Phi(\omega_s,\omega_i) $
defined in Eq.~(\ref{11}) characterize the emitted photon pair in
spectral domain, temporal two-photon amplitudes $
\tilde{\Phi}(t_s,t_i) $ defined as \cite{PerinaJr1999a}
\begin{equation}  
 \tilde{\Phi}(t_s,t_i) = \langle {\rm vac}| \hat{E}_s^{(+)}(0,t_s)
 \hat{E}_i^{(+)}(0,t_i) |\psi\rangle
\label{18}
\end{equation}
are useful for the description of photon pairs in time domain. The
substitution of Eqs.~(\ref{3}) and (\ref{10}) into Eq.~(\ref{18})
gives us the formula valid outside the fiber:
\begin{eqnarray} 
 \tilde{\Phi}_{\eta_s\eta_i}^{\eta_p}(t_s,t_i) &=& \frac{\hbar}{4\pi \varepsilon_0 c}
  \int d\omega_s \int d\omega_i \, \frac{\sqrt{\omega_s \omega_i}}
  {\sqrt{\bar{n}_{s,\eta_s}\bar{n}_{i,\eta_i}}} \,
  \nonumber \\
 & & \hspace{-15mm} \mbox{} \times
   \Phi_{\eta_s\eta_i}^{\eta_p}(\omega_s,\omega_i)  \exp(-i\omega_s t_s)\exp(-i\omega_i t_i) .
\label{19}
\end{eqnarray}

Photon pairs generated in an individual nonlinear process $
(\eta_p,\eta_s,\eta_i) $ usually have a complex spectral structure
that can be revealed by the Schmidt decomposition of spectral
two-photon amplitude $ \Phi_{\eta_s\eta_i}^{\eta_p} $,
\begin{equation} 
 \Phi_{\eta_s\eta_i}^{\eta_p}(\omega_s,\omega_i) =
 \sum_{k=0}^{\infty}
  \lambda_{\omega,k} f_{s,k}(\omega_s)f_{i,k}(\omega_s) .
\label{20}
\end{equation}
In Eq.~(\ref{20}), functions $ f_{s,k} $ and $ f_{i,k} $ form a
Schmidt dual basis and eigenvalues $ \lambda_{\omega,k} $ give
coefficients of the decomposition. Provided that these
coefficients are properly normalized ($ \sum_{k=0}^{\infty}
\lambda_{\omega,k}^2 =1 $) they determine the Schmidt number $
K_\omega $ of independent modes needed in the description
\cite{Law2004Analysis},
\begin{equation}   
 K_\omega = \frac{1}{\sum_{k=0}^{\infty} \lambda_{\omega,k}^4 } .
\label{21}
\end{equation}

\section{OAM decomposition of modes in the transverse plane }

Vector modes in the transverse plane have in general a complex
structure that, however, has to accord with rotational symmetry of
the fiber. For this reason, it is useful to decompose their
azimuthal dependencies into eigenmodes of OAM operator $
\hat{L}(\theta) $, $ \hat{L}(\theta) = -i\hbar \partial
/(\partial\theta) $, that take the form of harmonic functions
\cite{Molina-Terriza2001oam}:
\begin{equation}  
 t_l(\theta) = \frac{1}{\sqrt{2\pi}} \exp(il\theta).
\label{22}
\end{equation}
Convenience of this decomposition is even emphasized when
nonlinear processes are taken into account as there occurs the
conservation law of OAM number $ l $ \cite{Osorio2008oam}. This
law immediately follows from the integration over azimuthal angle
$ \theta $ in the interaction Hamiltonian $ \hat{H}_{\rm int} $
written in Eq.~(\ref{1}). For the considered SPDC process, this
law is expressed as
\begin{equation} 
 l_p = l_s + l_i ,
\label{23}
\end{equation}
where the subscripts indicate the participating fields.

The electric-field modes $ {\bf e}_{\eta}(r,\theta,\omega) $
involved in the interactions are vectorial, but their longitudinal
components $ {\bf e}_{z,\eta}(r,\theta,\omega) $ are usually at
least one order of magnitude smaller compared to their transverse
components $ {\bf e}_{r,\eta}(r,\theta,\omega) $, $ {\bf
e}_{\theta,\eta}(r,\theta,\omega) $ or $ {\bf
e}_{x,\eta}(r,\theta,\omega) $, $ {\bf
e}_{y,\eta}(r,\theta,\omega) $ \cite{Yeh2010waveguides,Snyder1983}. For this reason, we concentrate
our attention to the analysis of transverse components. The
analysis of cartesian transverse components $ {\bf
e}_{x,\eta}(r,\theta,\omega) $ and $ {\bf
e}_{y,\eta}(r,\theta,\omega) $ is more useful as they can easily
be experimentally obtained using optical polarizers. Moreover, the
$ x $ and $ y $ components of electric-field amplitude $ {\bf
e}_{\eta}(r,\theta,\omega) $ of the circularly polarized modes
given in Eq.~(\ref{8}) are only mutually shifted in azimuthal
variable $ \theta $ by $ \pi/2 $. That is why, we further pay
attention only to the $ x $ component $ {\bf
e}_{x,\eta}(r,\theta,\omega) \equiv e_{\eta}(r,\theta,\omega) $.

The mode functions $ e_{\eta}(r,\theta,\omega) $ depend in general
on three variables $ r $, $ \theta $ and frequency $\omega$.
Following the rules of quantum mechanics, the probability $ p $ of
detecting a photon in an OAM eigenstate $ t_l $ is given by the
formula \cite{Molina-Terriza2001oam}:
\begin{equation}   
 p_{l,\eta}(\omega) = \int rdr \left| \int d\theta \, t_l^*(\theta)
  e_{\eta}(r,\theta,\omega)\right|^2
\label{24}
\end{equation}
that expresses averaging over the radial variable $ r $.

As entangled photon pairs in their general form (for
hyper-entangled photons, see
\cite{Barreiro2005Generation,Kang2014Hyperentangled}) are emitted,
two-photon amplitudes $ \Phi $ depending on both transverse-plane
variables and frequencies are needed in their description. They
generalize the two-photon spectral amplitudes $
\Phi_{\eta_s\eta_i}^{\eta_p}(\omega_s,\omega_i) $ defined in
Eq.~(\ref{11}). In the usually considered spectral ranges several
nm wide, the two-photon amplitude $
\Phi(r_s,\theta_s,\omega_s,r_i,\theta_i,\omega_i) $ can be
approximately written in the following factorized form:
\begin{equation}   
 \Phi(r_s,\theta_s,\omega_s,r_i,\theta_i,\omega_i) \approx
  \Phi_{r\theta}(r_s,\theta_s,r_i,\theta_i) \,
  \Phi_{\omega}(\omega_s,\omega_i).
\label{25}
\end{equation}
The transverse part $ \Phi_{r\theta} $ of two-photon amplitude
can in principle be decomposed similarly as the spectral part $
\Phi_{\omega} $ in Eq.~(\ref{18}), i.e.
\begin{equation}   
 \Phi_{r\theta}(r_s,\theta_s,r_i,\theta_i) = \sum_{m}
  \lambda_{r\theta,m} \, g_{s,m}(r_s,\theta_s)g_{i,m}(r_i,\theta_i)
\label{26}
\end{equation}
using eigenvalues $ \lambda_{r\theta,m} $ and eigenfunctions $
g_{s,m} $ and $ g_{i,m} $. The eigenvalues $ \lambda_{r\theta,m} $
then determine the Schmidt number $ K_{r\theta} $ of independent
modes by the formula (\ref{21}). However, the decomposition
(\ref{26}) is usually difficult to achieve. Nevertheless, the
two-photon amplitude $ \Phi_{r\theta}(r_s,\theta_s,r_i,\theta_i) $
nearly factorizes into its radial and azimuthal parts due to a
simple radial dependence in our case. Then we can obtain an
approximate number $ K_\theta $ of modes from singular values $
\lambda_{\theta,l} $ of matrix $ {\rm F}_\theta $ defined as
\begin{eqnarray}  
 {\rm F}_{\theta,l_sl_i} &=& \Biggl[ \int r_sdr_s \int r_idr_i
  \Biggl| \int d\theta_s \int d\theta_i \nonumber \\
 & & t_{l_s}^*(\theta_s)
  t_{l_i}^*(\theta_i) \Phi_{r\theta}(r_s,\theta_s,r_i,\theta_i)
  \Biggr|^2 \Biggr]^{1/2}
\label{27}
\end{eqnarray}
using Eq.~(\ref{21}).

\section{Guided modes of a ring fiber}

We consider the generation of photon pairs around the wavelengths
$ \lambda_s^0 $ and $ \lambda_i^0 $ equal to $ 1.55~\mu$m using
the pump field at the wavelength $ \lambda_p^0 = 0.775~\mu $m.
>From the considerations of fields' propagation stability and
efficiency of the nonlinear interaction, the fiber was designed to
guide radial fundamental modes for wavelengths longer than
$1.2\mu$m. This can be assured by a suitable choice of geometry of
the fiber. It holds that the fundamental mode arises as the first
solution of dispersion equation with the highest value of
propagation constant $ \beta(\omega) $ and occurs even at the
lowest possible guided frequencies. Higher-order modes given by
subsequent solutions of the dispersion equation exist in general
only for higher frequencies $ \omega $. The higher the radial mode
number, the higher the threshold frequency $ \omega $. This
property allows us to exclude higher-order modes for the chosen
frequencies (wavelengths) by a suitable choice of radii $ r_1 $
and $ r_2 $ of the fiber ring core (Fig.~\ref{fig1}). Detailed
numerical calculations have revealed that the analyzed fiber with
its core extending from $ r_1 = 4~\mu$m to $ r_2 = 5.5~\mu$m
admits only the radial fundamental modes for the wavelengths
longer than $ 1.2~\mu$m.

\begin{figure}  
 \centering
 \subfigure[]{\includegraphics{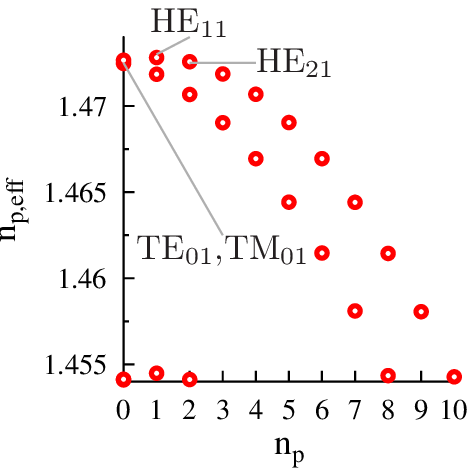}}
 \subfigure[]{\includegraphics{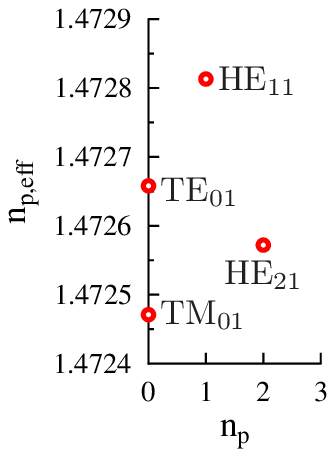}}
 \caption{(a) Effective refractive index $ n_{p,\rm eff} $ of the pump field
  in dependence on azimuthal number $ n_p $ for
  $ \lambda_{p}^0=0.775~\mu$m. In (b), detail of the graph around
  $ n_p = 0 $ is shown.}
\label{fig2}
\end{figure}

Effective indices of refraction $ n_{\rm p, eff} $ ($ n_{\rm p,
eff} = c\beta_p/\omega_p $) for the pump field at the wavelength $
\lambda_p^0 = 0.775~\mu $m are shown in Fig.~\ref{fig2}. They can
be indexed by two numbers, the first counts azimuthal modes and
the second radial modes in cylindrical coordinates. The higher the
value of the index number, the more complex the mode transverse
profile. Modes with the simplest transverse profiles are
interesting for the nonlinear interaction as they propagate with
low distortions and also allow to reach the greatest values of the
interaction overlap integral written in Eq.~(\ref{12}). From this
point of view, TE$ _{01} $, TM$ _{01} $, HE$ _{11} $, and HE$
_{21} $ modes with the greatest effective indices of refraction $
n_{\rm p, eff} $ are important (see Fig.~\ref{fig2}). Whereas
transverse components of TE$ _{01} $ and TM$ _{01} $ modes have a
complex structure from the point of view of OAM eigenmodes $
t_l(\theta) $ given by Eq.~(\ref{22}), transverse components of
modes HE$ _{11,R} $ and HE$ _{11,L} $ are close to eigenmode $
t_0(\theta) $. Transverse components of mode HE$ _{21,R} $ [HE$
_{21,L} $] are close to eigenmode $ t_{+1}(\theta) $ [$
t_{-1}(\theta) $] and so bear a nonzero OAM (for details, see
Fig.~\ref{fig4} below).

\begin{figure}  
 \centering
 \subfigure[]{\includegraphics{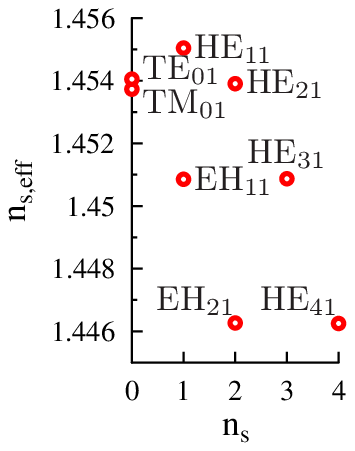}}
 \subfigure[]{\includegraphics{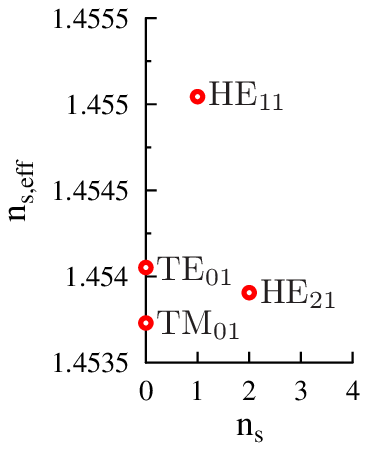}}
 \caption{(a) Effective refractive index $ n_{s,\rm eff} $ of the signal field
  in dependence on azimuthal number $ n_s $ for
  $ \lambda_{s}^0=1.55~\mu$m. In (b), detail of the graph around
  $ n_s = 0 $ is shown.}
\label{fig3}
\end{figure}

The signal and idler fields analyzed at the wavelength $
\lambda_s^0 = \lambda_i^0 = 1.55~\mu $m contain only radial
fundamental modes which effective indices of refraction $ n_{s,\rm
eff} $ are plotted in Fig.~(\ref{fig3}). In total 14 modes occur
in the analyzed spectral region: TE$ _{01} $ and TM$ _{01} $ modes
without a defined OAM and both left- and right-handed circularly
polarized modes HE$ _{11} $ ($ l= 0 $), HE$ _{21} $ ($ l= \pm 1
$), HE$ _{31} $ ($ l= \pm 2 $), HE$ _{41} $ ($ l= \pm 3 $), EH$
_{11} $ ($ l= \pm 2 $) and EH$ _{21} $ ($ l= \pm 3 $).

Profiles of the $ x $ and $ z $ components of signal
electric-field amplitudes $ {\bf e}_{\eta}(r,\theta) $ for four
simplest modes, TE$ _{01} $, TM$ _{01} $, HE$ _{11} $, and HE$
_{21} $, are shown in Fig.~\ref{fig4}. The $ y $ components of
electric-field amplitudes $ {\bf e}_{\eta}(r,\theta) $ have the
same amplitudes as the $ x $ components of $ {\bf
e}_{\eta}(r,\theta) $ but their phases are shifted by $ \pi/2 $
with respect to the phases of the $ x $ components. We note that
the pump modes have similar profiles as the signal modes, they are
only more localized inside the core ring as a consequence of their
half wavelength relative to the signal one.
\begin{figure}  
 \centering
 \includegraphics{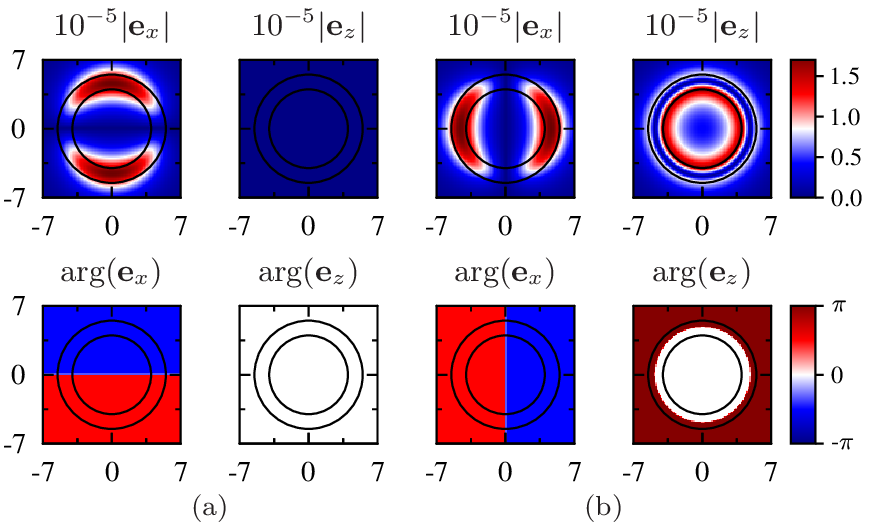}
 \includegraphics{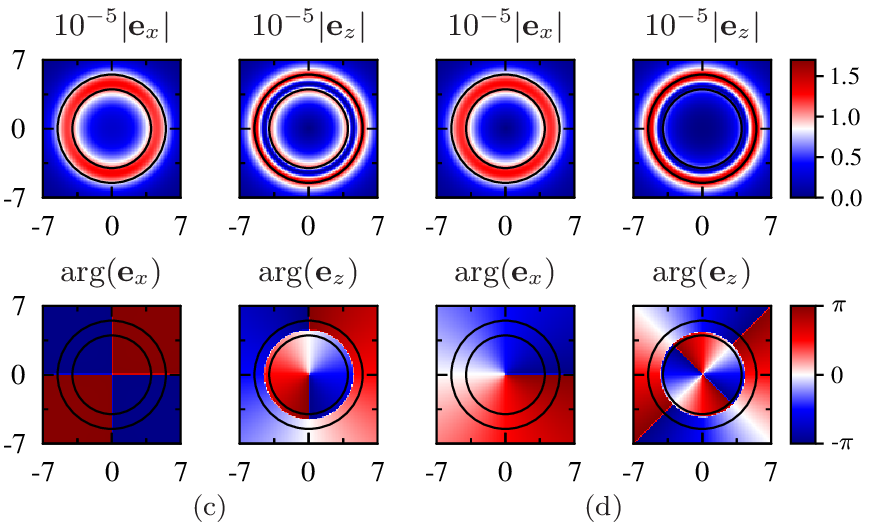}
 \caption{[Color online:]
  Amplitude and phase of components $ {\bf e}_{x}(x,y) $ and $ {\bf e}_{z}(x,y) $
  of electric-field amplitudes for modes TE$ _{01} $ (a), TM$ _{01} $ (b),
  HE$ _{11,R} $ (c), and HE$ _{21,R} $ (d) for the signal field at
  $ \lambda_{s}^0=1.55~\mu$m; $ x = r \cos(\theta) $, $ x = r \sin(\theta) $.
  The cartesian $ x $ and $ y $ axes' labels are in $ \mu $m and
  the components are normalized according to $ \int dxdy |{\bf e}_{x,z}(x,y)|^2 = 1 $.}
\label{fig4}
\end{figure}

\begin{figure}  
 \centering
 \includegraphics{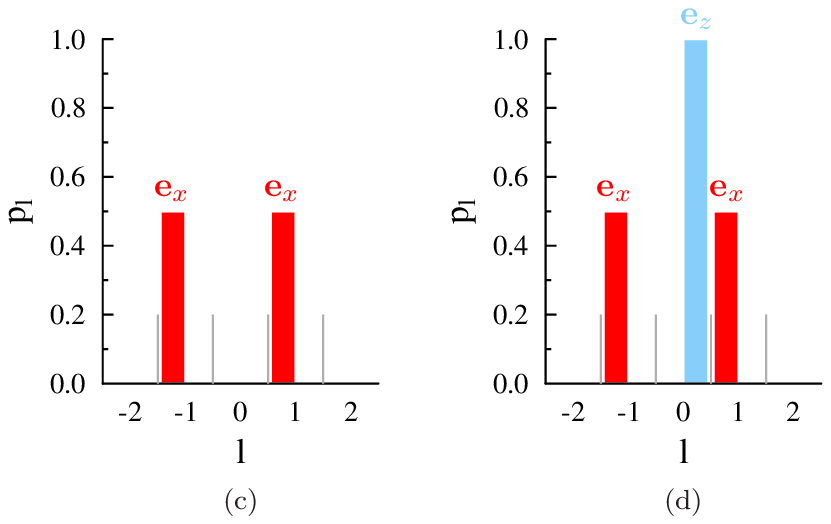}
 \includegraphics{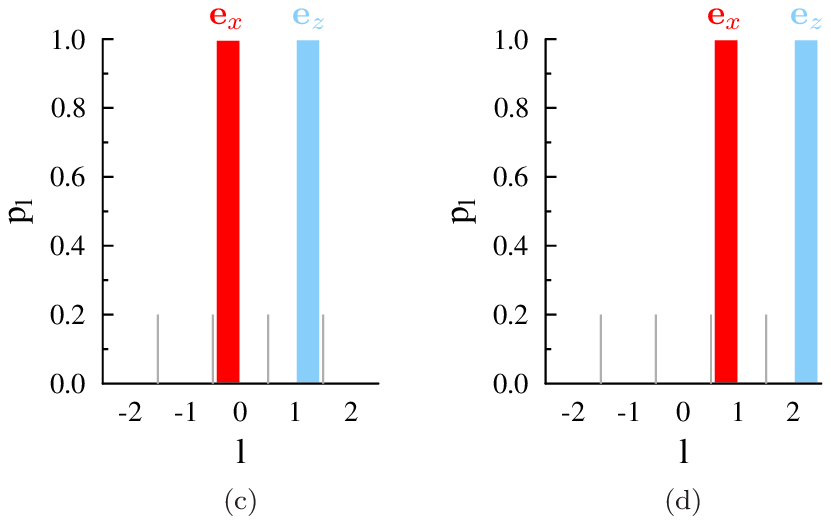}

 \caption{[Color online:] Probabilities $ p_l $ of measuring an OAM eigenmode $ t_l $
  for the $ x $ and $ z $ components of electric-field amplitude
  $ {\bf e}_{\eta}(r,\theta,\omega) $
  for modes TE$ _{01} $ (a), TM$ _{01} $ (b),
  HE$ _{11,R} $ (c), and HE$ _{21,R} $ (d) for the signal field at
  $ \lambda_{s}^0=1.55~\mu$m.}
\label{fig5}
\end{figure}

The weights of individual OAM eigenmodes in the above modes
determined by probabilities $ p $ in Eq.~(\ref{24}) are important
for judging efficiency of the nonlinear interaction as it obeys
the conservation law of OAM expressed in Eq.~(\ref{23}). The
probabilities $ p $ determined for the most useful modes TE$ _{01}
$, TM$ _{01} $, HE$ _{11} $, and HE$ _{21} $ of the signal field
are depicted in Fig.~\ref{fig5}. Whereas several OAM eigenmodes
are essential for building TE$ _{01} $ and TM$ _{01} $ modes, the
OAM eigenmode $ t_{0}(\theta) $ [$ t_{+1}(\theta) $ and $
t_{-1}(\theta) $] dominates in the $ x $ and $ y $ components of
electric-field amplitude $ {\bf e}_{\eta}(r,\theta) $ of modes HE$
_{11,R} $ and HE$ _{11,L} $ [HE$ _{21,R} $ and HE$ _{21,L} $]. On
the other hand, the $ z $ components of electric-field amplitudes
$ {\bf e}_{\eta}(r,\theta) $ usually contain OAM eigenmodes $ t_l
$ with $ l $ in absolute value greater by one compared to their $
x $ and $ y $ components. So the component $ {\bf
e}_{z,\mathrm{HE}_{m1,R}}(r,\theta) $ [$ {\bf
e}_{z,\mathrm{HE}_{m1,L}}(r,\theta) $] is formed by OAM eigenstate
$ t_{+m}(\theta) $ [$ t_{-m}(\theta) $] for $ m=1,2 $.

These modes of the pump, signal and idler fields can be combined
in several different ways in order to arrive at an efficient
nonlinear interaction among individual modes. This interaction is
efficient provided that the conservation laws of energy and OAM
 together with quasi-phase-matching are fulfilled. Period
$ \Lambda $ of periodical poling is the only free parameter that
allows us to choose among several individual nonlinear processes.
In the following three sections, we analyze different processes
that give us both spectrally narrow- and broad-band photon pairs
as well as photon pairs entangled in OAM numbers.

\section{Generation of photon pairs with nonzero OAM numbers}

Pump modes with zero OAM numbers $ l_p $ are suitable for the
generation of spectrally broad-band photon pairs whereas pump
modes with non-zero OAM are optimal for the emission of spectrally
narrow-band photon pairs. When the conservation law of OAM in
Eq.~(\ref{23}) is applied to pump modes HE$_{11,R}$ and
HE$_{11,L}$ with $ l_p = 0 $, the signal $ l_s $ and idler $ l_i $
OAM numbers have to have the same absolute value. The signal and
idler modes then naturally have similar properties, which allow
for a broad-band photon-pair generation (see the next section). On
the other hand, if the pump beam is in mode HE$_{21,R}$ with $ l_p
= +1 $ (or its left-handed circularly polarized variant
HE$_{21,L}$ with $ l_p = -1 $) the conservation law of OAM
suggests the signal and idler modes with different OAMs. The
down-converted modes are then expected to have different
properties and the emission of photon-pairs is assumed to be
narrow-band and non-degenerate. Stability of the pump mode
HE$_{21,R}$ follows from the graph in Fig.~\ref{fig2} that
identifies modes TE$_{01}$ and TM$_{01}$ as the closest modes with
respect to effective refractive index $ n_{p,\rm eff} $. However,
differences $ \Delta n_{p,\rm eff} $ between the modes ($ \Delta
n_{p,\rm eff} = -9\times 10^{-5} $ for mode TE$_{01}$, $ \Delta
n_{p,\rm eff} = 1\times 10^{-4} $ for mode TM$_{01}$) are high
enough to guarantee stable guiding of mode HE$_{21}$ without
crosstalk.

The signal and idler modes fulfilling the conservation of OAM
together with the pump HE$_{21,R}$ mode are summarized in
Tab.~\ref{tab1}. However, only the variants with the signal HE$
_{21,R} $ mode and idler HE$ _{11,R} $ and HE$ _{11,L} $ modes are
sufficiently stable. The fundamental modes HE$ _{11,R} $ and HE$
_{11,L} $ are the most stable. In detail, the difference $ \Delta
n_{s,\rm eff} $ of refraction indices of modes HE$_{11}$ and the
closest mode TE$_{01}$ equals $ 1\times 10^{-3} $, whereas $
\Delta n_{s,\rm eff} $ for mode HE$_{21}$ and the closest mode
TE$_{01}$ is $ 1.5\times 10^{-4} $.
\begin{table}  
 \begin{center}
  \begin{tabular}{|S|S|S|S|}
  \hline
  pump & \multicolumn{3}{|c|}{HE$_{21,R}$} \\
  \hline
  $l_{p}$ & \multicolumn{3}{|c|}{+1} \\
  \hline
  signal & HE$_{21,R}$ & \bigcell{c}{HE$_{31,R}$\\EH$_{11,R}$} & \bigcell{c}{HE$_{41,R}$\\EH$_{21,R}$} \\
  \hline
  $l_{s}$ & 1 & 2 & 3\\
  \hline
  idler & \bigcell{c}{HE$_{11,R}$\\HE$_{11,L}$} & HE$_{21,L}$ & \bigcell{c}{HE$_{31,L}$\\ EH$_{11,L}$} \\
  \hline
  $l_{s}$ & 0 & -1 & -2 \\
  \hline
  \end{tabular}
 \end{center}
 \caption{Possible combinations of pump, signal and idler modes with their
 OAM numbers $ l $ (in the weakly-guiding approximation \cite{Snyder1983})
 fulfilling the conservation of OAM.}
\label{tab1}
\end{table}

\begin{figure}  
\includegraphics[scale=1]{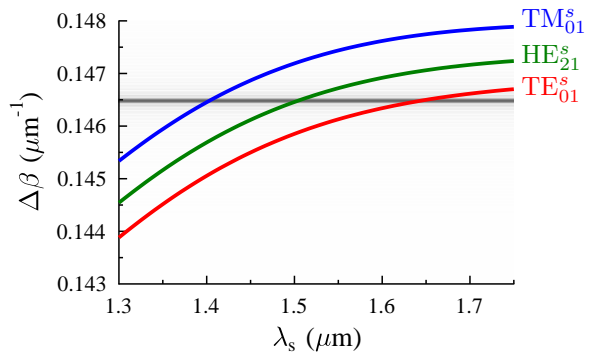}
 \caption{[Color online:] Phase mismatch $\Delta \beta $ for nonlinear processes
  (HE$_{21}^p$,HE$_{21}^s$,HE$_{11}^i$),
  (HE$_{21}^p$,TE$_{01}^s$,HE$_{11}^i$) and
  (HE$_{21}^p$,TM$_{01}^s$,HE$_{11}^i$).
  The gray horizontal pattern describes spatial spectrum $ \tilde{\chi}^{(2)}(\beta) $
  of a rectangular QPM grating with $ \Lambda = 42.9~\mu$m; $ L=10
  $~cm.}
\label{fig6}
\end{figure}

Also the signal TE$ _{01} $ and TM$ _{01} $ modes may participate
in the nonlinear interaction as they are partially composed of OAM
eigenmodes with $ l_s = +1 $ (see Fig.~\ref{fig6}). However, these
modes are not suitable for transmission of photons as they do not
have a well defined OAM. They can be spectrally separated from the
combinations of modes discussed above owing to different
propagation constants. They lead to different values of nonlinear
phase mismatch $ \Delta\beta $ for the considered individual
nonlinear processes (HE$_{21}^p$,HE$_{21}^s$,HE$_{11}^i$),
(HE$_{21}^p$,TE$_{01}^s$,HE$_{11}^i$) and
(HE$_{21}^p$,TM$_{01}^s$,HE$_{11}^i$). The dependence of nonlinear
phase mismatch $ \Delta\beta $ on signal wavelength $ \lambda_s $
for cw pumping plotted in Fig.~\ref{6} shows that a sufficiently
narrow spatial spectrum $ \tilde{\chi}^{(2)} $ of QPM grating [see
Eq.~(\ref{14})] provides spectral separation of different
nonlinear processes. Width $ \Delta\tilde{\chi}^{(2)} $ of spatial
spectrum can easily be varied by the length $ L $ of the grating.
The longer the grating, the narrower the spectrum $
\tilde{\chi}^{(2)} $ and also the narrower the signal- and
idler-field spectra. Individual nonlinear processes are thus
better separated for longer QPM gratings. Therefore a suitable
length of the grating has to be found. A 10-cm long rectangular
grating with period $ \Lambda = 42.9~\mu$m available by a simple
fabrication method \cite{Zhang2007Demonstration} (see
Fig.~\ref{fig6} for its spectrum $ \tilde{\chi}^{(2)} $) satisfies
the requirement. It allows the generation of signal photons around
the wavelength $ \lambda_s^0=1.5~\mu$m accompanied by idler
photons around the wavelength $ \lambda_i^0=1.6~\mu$m in the
nonlinear process (HE$_{21}^p$,HE$_{21}^s$,HE$_{11}^i$). Intensity
spectral width $ \Delta\tilde{\chi}^{(2)} $ equals $ 2\times
10^{-4}~\mu$m$^{-1}$ (full width at half maximum, FWHM) for this
grating and guarantees the amount of unwanted photons at the level
of 1\%.

\begin{figure}  
 \includegraphics[scale=1]{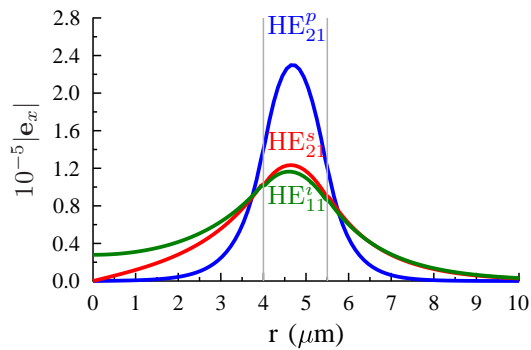}
 \caption{[Color online:] Absolute value $ |{\bf e}_{x}| $ of the $ x $ component of
  electric-field amplitude depending on radius $ r $ for pump mode
  HE$_{21}^p$, signal mode HE$_{21}^s$ and idler mode
  HE$_{11}^i$. Normalization is such that
  $ \int dxdy \,  |{\bf e}_{x}(x,y)|^2 = 1 $.}
\label{fig7}
\end{figure}

The number of generated photon pairs depends on the overlap
integral containing the product of pump, signal and idler
electric-field amplitudes in the transverse plane [see
Eq.~(\ref{12})]. The value of this integral in the azimuthal angle
is maximized due to the conservation of OAM. The maximal available
value of this integral then depends on radial mode profiles that
are shown in Fig.~\ref{fig7} for the chosen nonlinear process. It
holds in general that the lower the number of minima in radial
intensity profiles, the greater the number of generated photon
pairs. This favors modes with lower mode numbers.

\begin{figure}   
 \centering
  \includegraphics[scale=1]{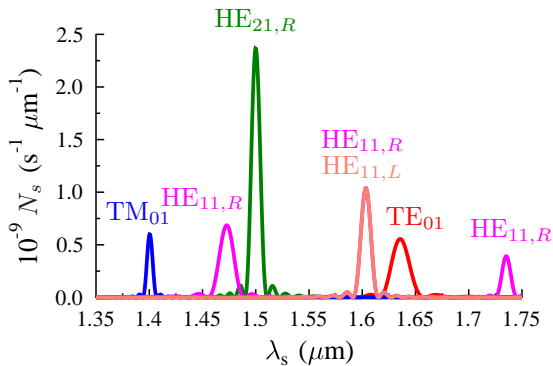}
  \caption{[Color online:] Spectral photon-number density $ N_s $ created by modes
   HE$_{21,R}$, HE$_{11,R}$, HE$_{11,L}$, TE$_{01}$, and TM$_{01}$ in dependence on
   wavelength $ \lambda_s $; $ N_s(\omega_s)=\sum_{\eta_p,\eta_s,\eta_i}
   N_{s,\eta_s,\eta_i}^{\eta_p}(\omega_s)$}
\label{fig8}
\end{figure}

Six well separated peaks occur in the down-converted field
spectrum $ N_s(\lambda_s) $ shown in Fig.~\ref{fig8}. The most
intensive peak at $ \lambda_s = 1.5~\mu$m belongs to mode
HE$_{21,R}^s$ and originates in the nonlinear processes
(HE$_{21,R}^p$,HE$_{21,R}^s$,HE$_{11,L}^i$) and
(HE$_{21,R}^p$,HE$_{21,R}^s$,HE$_{11,R}^i$). The accompanying
peaks at $ \lambda_i = 1.603~\mu$m correspond to modes
HE$_{11,L}^i$ and HE$_{11,R}^i$ with the same weight. The curves
in Fig.~\ref{fig8} confirm that these desired peaks can be well
separated by frequency filters from the unwanted ones. We note
that the modes HE$_{11,L}^i$ and HE$_{11,R}^i$ with the same
spectra cannot be separated and in fact form a common quantum
superposition state. The efficiency of spectral separation in ring
fibers is similar to that found in nonlinear waveguides with SPDC
\cite{Karpinski2009,Machulka2013}. Spectral width of the peak at $
\lambda_s = 1.5~\mu$m equals 9.41~nm (FWHM). The peak occurring at
$ \lambda_s = 1.4~\mu$m belongs to TM$ _{01}$ mode and forms a
pair together with the peak at $ \lambda_i = 1.73~\mu$m given by
mode HE$_{11,R}^i$. Mode TE$ _{01}^s$ is responsible for the peak
at $ \lambda_s = 1.63~\mu$m that occurs together with the peak at
$ \lambda_i = 1.47~\mu$m established by mode HE$_{11,R}^i$. We
note that small oscillations at the wings of the peaks reflect the
shape of spatial spectrum $ \tilde{\chi}^{(2)}(\beta) $ of QPM
grating.

As follows from Fig.~\ref{fig8}, photon-pair density $ N_s $
attains its maximum value at $ 2.4 \times
10^9~\mbox{n}\mathrm{m}^{-1}\mbox{s}^{-1} $ for 1~W of the pump
power. Taking into account the peak spectral width, around 20
photon pairs per 1~s and $ 1~\mu $W of pumping are expected in
modes (HE$_{21,R}^p$,HE$_{21,R}^s$,HE$_{11}^i$) provided that
appropriate spectral filters are used. The number of generated
photon pairs can be increased by considering longer fibers. It can
be shown theoretically that the number of photon pairs increases
better than linearly with the fiber length. Also narrowing of the
emitted spectra occurs with the increasing fiber length. On the
other hand, fabrication imperfections as well as non-ideal
alignment of the nonlinear interaction in the laboratory reduces
these numbers by one or two orders in magnitude
\cite{Zhu2012direct}.

Photon pairs are emitted in states entangled in signal and idler
frequencies due to the conservation law of energy. This results in
sharp temporal correlations in detection times of the signal and
idler photons. For the spectra approx. 10~nm wide, typical
entanglement times quantifying these correlations are in hundreds
of fs (for details, see Fig.~\ref{fig11} below)
\cite{PerinaJr2011pbg}.

\section{Generation of spectrally broad-band photon pairs}

\begin{figure}  
\includegraphics[scale=1]{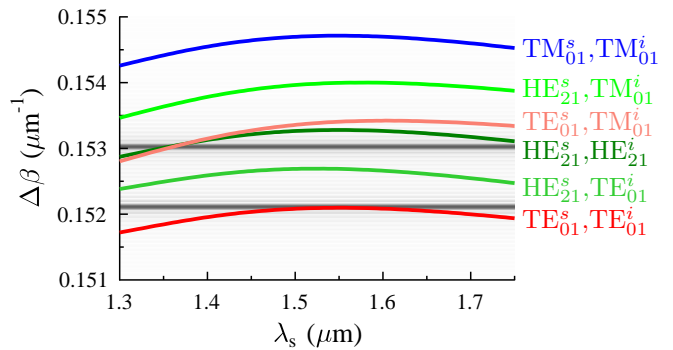}
 \caption{[Color online:] Phase mismatch $\Delta \beta $ for nonlinear processes
  pumped by mode HE$_{11,R}^p$ with signal and idler fields in
  modes HE$_{21}$, TE$_{01}$ and TM$_{01}$ in dependence on wavelength of
  signal photon $ \lambda_s $. The gray horizontal pattern
  describes spatial spectra $ \tilde{\chi}^{(2)}(\Delta \beta) $
  of rectangular nonlinear modulation with $ \Lambda = 41.06~\mu$m (upper pattern) and
  $ \Lambda = 42.28~\mu$m (lower pattern); $ L=10 $~cm.}
\label{fig9}
\end{figure}

As it has already been discussed above, the pump field in a
HE$_{11,R}$ (or HE$_{11,L}$) mode with $ l_p = 0 $ allows to
generate spectrally broad-band photon pairs achievable usually in
chirped poled nonlinear materials
\cite{Svozilik2009qpm,Svozilik2010qpm}. This is a consequence of
flat spectral dependencies of phase mismatches $ \Delta\beta $ of
individual nonlinear processes conserving OAM (see
Fig.~\ref{fig9}). Stable down-converted modes of LP$ _{11} $ family, HE$_{21} $,
TE$_{01}$ and TM$_{01}$, can take part in this interaction. The
curves in Fig.~\ref{fig9} indicate that the nonlinear processes
(HE$_{11,R}^p$,HE$_{21,R}^s$,HE$_{21,L}^i$),
(HE$_{11,R}^p$,HE$_{21,L}^s$,HE$_{21,R}^i$),
(HE$_{11,R}^p$,TE$_{01}^s$,TM$_{01}^i$), and
(HE$_{11,R}^p$,TM$_{01}^s$,TE$_{01}^i$) occur nearly
simultaneously and thus may provide a more complex state. On the
other hand, the processes (HE$_{11,R}^p$,TE$_{01}^s$,TE$_{01}^i$)
and (HE$_{11,R}^p$,TM$_{01}^s$,TM$_{01}^i$) can easily be
separated from other processes for sufficiently narrow spatial
spectra $ \tilde{\chi}^{(2)}(\beta) $, similarly as in the case discussed
in Sec.~V.

\begin{figure}   
 \centering
  \includegraphics[scale=1]{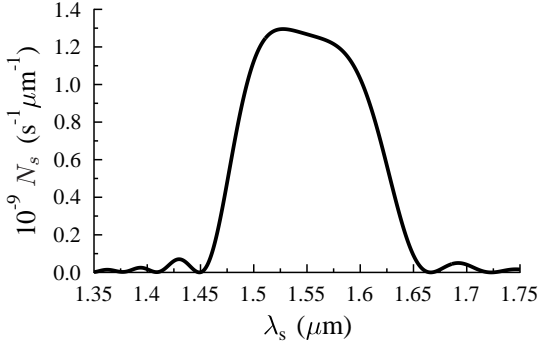}
  \caption{Spectral photon-number density $ N_s $ originating
  in nonlinear process (HE$_{11,R}^p$,TE$_{01}^s$,TE$_{01}^i$); $ \Lambda =
  42.28~\mu$m, $ L=10$~cm in dependence on wavelength $ \lambda_s $.}
\label{fig10}
\end{figure}

\begin{figure}   
 \centering
   \includegraphics[scale=1]{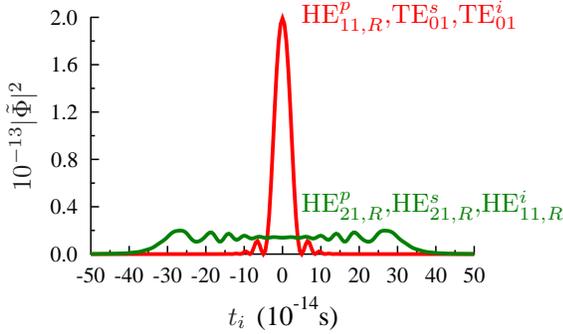}
  \caption{[Color online:] Probability density $ p_{t,i} $, $ p_{t,i} = {\cal C}|\tilde{\Phi}(0,t_i)|^2 $,
   as a function of idler-photon detection time $ t_i $
   for a signal photon detected at time $ t_s=0$~s for processes
   (HE$_{21,R}^p$,HE$_{21,R}^s$,HE$_{11,R}^i$) ($ \Lambda =
   42.9~\mu$m) and (HE$_{11,R}^p$,TE$_{01}^s$,TE$_{01}^i$) ($ \Lambda =
   42.28~\mu$m), $ L=10$~cm. Constant $ {\cal C} $ is defined such
   that $ \int_{-\infty}^{\infty} dt_i p_{t,i}(t_i)=1 $.}
\label{fig11}
\end{figure}

As an example, we consider the nonlinear interaction with TE$_{01}
$ signal and idler modes. This interaction is achieved for period
$ \Lambda $ of the nonlinear modulation equal to $ 42.28~\mu$m.
Signal photon-number density $ N_s(\lambda_s) $ for this process
and 10-cm long QPM grating attains its maximum at degenerate
wavelength $ \lambda_s^0 = 1.55~\mu$m where a 142-nm wide peak
occurs (FWHM, see Fig.~\ref{fig10}). Around 150 photon pairs per
1~s and $ 1~\mu $W of pumping are emitted in this process. The
obtained spectrum is approx. 15 times broader compared to that of
the process analyzed in Sec.~V. This implies considerably sharper
temporal features of photon pairs generated by the process
(HE$_{11,R}^p$,TE$_{01}^s$,TE$_{01}^i$). Profiles of probability
densities $ p_{t,i} $ of detecting an idler photon at time $ t_i $
conditioned by detection of a signal photon at time $ t_s =0$~s
for both cases are compared in Fig.~\ref{fig11} confirming this
fact. Whereas the probability-density width equals
$4.5\times10^{-14}$~s (FWHM) for the spectrally broad-band process
(HE$_{11,R}^p$,TE$_{01}^s$,TE$_{01}^i$), it attains
$63.5\times10^{-14}$~s for the spectrally narrow-band process
(HE$_{21,R}^p$,HE$_{21,R}^s$,HE$_{11,R}^i$). We note that sharp
temporal correlations are important in metrology as they determine
the available temporal resolution
\cite{Carrasco2004,Fraine2012dispersion}.

\section{Generation of photon pairs entangled in OAM numbers}

\begin{figure}   
 \centering
  \includegraphics[scale=1]{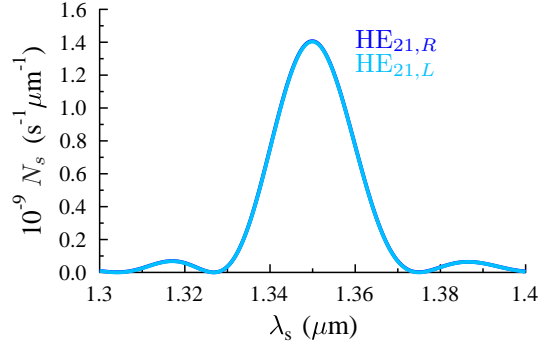}
  \caption{[Color online:] Spectral photon-number density $ N_s $ arising
  from nonlinear processes (HE$_{11,R}^p$,HE$_{21,R}^s$,HE$_{21,L}^i$) and
  (HE$_{11,R}^p$,HE$_{21,L}^s$,HE$_{21,R}^i$) in dependence on wavelength
  $ \lambda_s $. The curves nearly coincide;
  $ \Lambda = 41.06~\mu$m, $ L=10$~cm.}
\label{fig12}
\end{figure}

Pumping the fiber with a HE$_{11,R} $ (or HE$_{11,L} $) mode is
interesting even in the case when more LP$_{11} $ modes
participate in the nonlinear interaction. Period $ \Lambda $ of
nonlinear modulation equal to $ 41.06~\mu$m  provides suitable
conditions for four nonlinear processes
(HE$_{11,R}^p$,HE$_{21,R}^s$,HE$_{21,L}^i$),
(HE$_{11,R}^p$,HE$_{21,L}^s$,HE$_{21,R}^i$),
(HE$_{21,R}^p$,TE$_{01}^s$,TM$_{01}^i$) and
(HE$_{21,R}^p$,TM$_{01}^s$,TE$_{01}^i$) (see Fig.~\ref{fig9}). The
last two processes do not contribute to photon-pair generation as
they have zero overlap integrals given in Eq.~(\ref{12}). In the
first two nonlinear interactions, the signal and idler photons are
generated with OAM numbers equal to $ \pm 1 $ and $ \mp 1 $. State
$ |\psi_{l_s,l_i} \rangle $ entangled in OAM numbers
\cite{Svozilik2012entanglement} [$ | \psi_{l_s,l_i} \rangle = C_1
| l_s=1 \rangle_s | l_i= -1 \rangle_i + C_2 | l_s=-1 \rangle_s |
l_i=1 \rangle_i $, $C_1$ and $C_2$ are constants] can thus be
obtained at wavelengths $ \lambda_s = 1.35~\mu$m and $ \lambda_i =
1.82~\mu$m. As both processes have nearly equal intensities, a
generated state close to the maximally entangled state is
expected. Also radial profiles of the emitted photons are close to
each other which justifies the use of formula (\ref{27}) for the
determination of Schmidt number $ K_{\theta} $. It gives $
K_{\theta} = 1.998 $. For comparison, the exact numerical
decomposition described in Eq.~(\ref{26}) provides $ K_{\theta} =
1.994 $. The obtained peak in the signal photon-number density $
N_s(\lambda_s) $ is 21~nm wide (FWHM) and its profile is shown in
Fig.~\ref{fig12}. The curve plotted in Fig.~\ref{fig12}
corresponds to 30 signal photons generated per 1~s and $ 1~\mu $W
of pumping, which characterizes an intense source of photon pairs.

The generated state is simultaneously entangled also in the signal
and idler frequencies and its state can be expressed as
\begin{eqnarray} 
 |\psi \rangle \hspace{-2mm}&=&  \hspace{-2mm} \int d\omega_s d\omega_i\, \Phi_{1_s,-1_i}(\omega_s,\omega_i) | l_s=1, \omega_s
  \rangle_s | l_i =-1,\omega_i \rangle_i \nonumber \\
 & & + \Phi_{-1_s,1_i}(\omega_s,\omega_i) |l_s= -1, \omega_s
 \rangle_s| l_i=1, \omega_i \rangle_i.
\label{28}
\end{eqnarray}
We analyze spectral entanglement assuming separability of the
spectral profile and that in the transverse plane for both fields.
We also analyze the two-photon spectral amplitude $
\Phi_{1_s,-1_i}(\omega_s,\omega_i) $ arising from the process
(HE$_{11,R}^p$,HE$_{21,R}^s$,HE$_{21,L}^i$) and note that the
two-photon amplitude $ \Phi_{-1_s,1_i}(\omega_s,\omega_i) $ of
process (HE$_{11,R}^p$,HE$_{21,L}^s$,HE$_{21,R}^i$) is very
similar to the former one. As the amount of spectral entanglement
depends on the pump-field spectral width $ \sigma_p $, we consider
the Gaussian spectrum $ {\cal E}_p $ centered at frequency $
\omega_p^0 $ corresponding to $ \lambda_p^0 = 0.775~\mu $m,
\begin{equation}  
 {\cal E}_p(\omega) = \sqrt{\sqrt{\frac{2}{\pi}}\frac{1}{\sigma_p}}
  \exp\left[-\frac{(\omega-\omega_p^0)^2}{\sigma_p^2} \right] .
\label{29}
\end{equation}

\begin{figure}   
 \centering
   \includegraphics[scale=1]{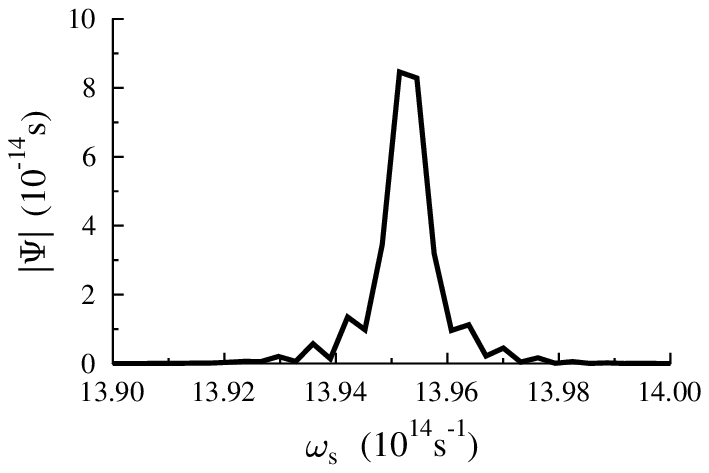}
  \includegraphics[scale=1]{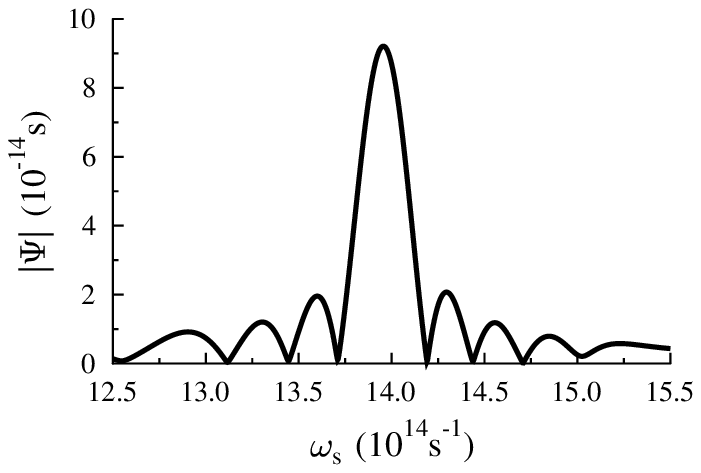}
  \caption{Cut of absolute value $ |\Phi_{1_s,-1_i}(\omega_s,\omega_i)| $ of two-photon spectral
   amplitude along the line (a) $ \omega_s = \omega_i - \omega_i^0+\omega_s^0 $ and (b) $
   \omega_s = \omega_p^0 - \omega_i $ appropriate for the process
   (HE$_{11,R}^p$,HE$_{21,R}^s$,HE$_{21,L}^i$) pumped by a pulsed
   field; $ \sigma_p =0.85 $~nm, $ \Lambda = 41.06~\mu$m, $ L=10$~cm.
   It holds that $ \int d\omega_s d\omega_i \, |\Phi(\omega_s,\omega_i)|^2 = 1 $.}
\label{fig13}
\end{figure}

The two-photon spectral amplitude $
\Phi_{1_s,-1_i}(\omega_s,\omega_i) $ considered for a pulsed pump
field has a typical elliptical shape with axes oriented at
directions $ \omega_s = \omega_i-\omega_i^0+\omega_s^0 $ and $
\omega_s = \omega_p^0 - \omega_i $. For the analyzed
configuration, the pump-field spectrum cannot be wider than $
\sigma_p = 0.85 $~nm (the corresponding intensity FWHM equals
2~nm) to assure negligible contributions from other nonlinear
processes discussed above. In this case, the two-photon amplitude
$ \Phi_{1_s,-1_i} $ is elongated along the direction $ \omega_s =
\omega_p^0 - \omega_i $. This is caused by the fact that the
extension of amplitude $ \Phi_{1_s,-1_i} $ in direction $ \omega_s
= \omega_i -\omega_i^0+\omega_s^0$ is limited by the product of
pump-field spectrum $ {\cal E}_p $ and spatial spectrum $
\tilde{\chi}^{(2)} $ of nonlinear modulation. As shown in
Fig.~\ref{fig13}(a) for the pump field with width $ \sigma_p =0.85
$~nm, spatial spectrum $ \tilde{\chi}^{(2)} $ introduces
oscillations in this direction. The extension of amplitude $
\Phi_{1_s,-1_i} $ in direction $ \omega_s = \omega_p^0 - \omega_i
$ depends on phase-matching properties of the structure as well as
on the pump-field spectrum. This admits much broader profiles, as
documented in Fig.~\ref{fig13}(b). Oscillations in spectrum $
\tilde{\chi}^{(2)} $ of nonlinear grating are also visible in this
profile and reflect profiles of dispersion curves.

\begin{figure}   
 \centering
  \includegraphics[scale=1]{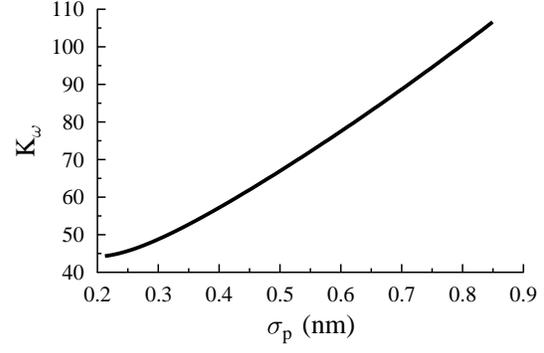}
  \caption{Number $ K_\omega $ of independent spectral modes as it depends
   on pump-field spectral width $ \sigma_p $; $ \Lambda = 41.06~\mu$m, $ L=10$~cm.}
\label{fig14}
\end{figure}

There typically occur several tens of independent spectral modes
for the considered pulsed pumping. The number $ K_\omega $ of
independent spectral modes determined by formula (\ref{21})
increases nearly linearly with the increasing pump-field spectral
width $ \sigma_p $ in the interval depicted in Fig.~\ref{fig14}.
This originates in considerable broadening of the signal- and
idler-field spectra with the increasing values of spectral width $
\sigma_p $. The overall number of independent modes is given by
the product $ K_\theta K_\omega $ of numbers of modes in the
spectral and azimuthal variables and thus reaches approx. 200 for
the pump field having 0.85-nm wide spectrum. All these modes can,
in principle, be used for quantum communications for delivering
entangled information.

We have considered a ring fiber 10~cm long as it can be fabricated
by a simple method \cite{Zhang2008poling}. However, there exists a
more sophisticated fabrication method allowing production of ring
fibers up to 1~m long \cite{Fokine2002}. The numbers of generated
photon pairs more than one order of magnitude greater are expected
in such fibers.

In many applications, the signal-to-noise ratio of a photon-pair
source is an important parameter. In the analyzed ring fiber, we
can identify three sources of noise. The first source is related
to the presence of other nonlinear processes. As the fiber has
losses, one photon from a generated photon pair can be lost
leaving the remaining photon in the form of noise. Finally, a
photon pair can be emitted into an unwanted pair of modes and so
both its photons contribute to the noise. However, it has been
shown in \cite{Zhu2012direct} that the Raman scattering as well as
other nonlinear processes are negligible in regular fibers with
the same material structure. As for the broken photon pairs, any
measurement based on the detection of photon coincidences
eliminates this kind of noise. Our results have shown that the
probability of generation of a photon pair into an unwanted pair
of modes is lower than 1/100 for the discussed configuration.
Thus, all three sources of noise can be neglected.

The discussed noise weakens entanglement of the generated state
entangled in OAM numbers. This weakening can be quantified, e.g.,
using the Clauser-Horne-Shimony-Holt (CHSH) form of the Bell
inequalities \cite{Clauser1969Proposed}. To simplify calculations,
we first determine a reduced statistical operator $\hat{\rho}_{\rm
OAM} $ corresponding to the state $ |\psi \rangle $ in
Eq.~(\ref{28}) reduced over the signal ($ \omega_s $) and idler ($
\omega_i $) frequencies. Considering additional noise with
relative weight $ p $, an appropriate statistical operator $
\hat{\rho}'_{\rm OAM} $ can be expressed as
\begin{equation} 
 \hat{\rho}'_{\rm OAM}= (1-p) \hat{\rho}_{\rm
  OAM}+p\frac{\hat{I}}{4}
\end{equation}
using the unity operator $ \hat{I} $. Maximal violation of the
CHSH inequalities occurs under conditions discussed in
\cite{Horodecki1995Violating}. In this case and assuming $ p=0.01
$, parameter $ S $ quantifying this violation ($ S > 2 $) equals
2.8. The boundary value of parameter S (S = 2) is observed for $ p
=0.283 $, which does not represent a real limitation for
experiments. For comparison, recent measurements with states
entangled in OAM numbers have reached $ S=2.78 $ for $l\pm1$
\cite{Mclaren2012Entangled} and $ S=2.69 $ for $l\pm2$
\cite{Leach2009Violation}.

We note that also vortex fibers have been considered
\cite{Bozinovic2013Terabit} for the propagation of optical fields
with nonzero OAM. Compared to ring fibers, they contain an
additional central core. As a consequence, their fundamental modes
HE$ _{11} $ are more stable. This advantage can be exploited also
when generating photon pairs as the analysis of SPDC in vortex
fibers is similar to that shown here for the ring fibers.

\section{Conclusions}

A nonlinear thermally-poled ring fiber has been considered as a
source of photon pairs arising in the process of spontaneous
parametric down-conversion. It has been shown that several
combinations of stable pump, signal and idler spatial modes of the
fiber are suitable for efficient photon-pair generation depending
on the period of $ \chi^{(2)} $ modulation introduced into the
fiber. Spectrally narrow-band photon pairs in OAM eigenstates
emitted in spectrally separated regions can be achieved this way.
Also broad-band photon pairs with spectra more than 100~nm wide
can be obtained in the fiber. Even photon-pair states entangled in
OAM eigenstates can be generated. For a pulsed pump field with
2-nm wide spectrum (FWHM), combined spectral and azimuthal
effective dimensions of the emitted entangled states can reach
200. The considered 10-cm long fiber allows to generate these
states with photon-pair fluxes reaching hundreds of pairs per
second and $ \mu $W of pumping. Higher photon-pair fluxes can be
obtained from longer fibers. These results show that nonlinear
ring and vortex fibers can be designed such that they can emit
intensive entangled photon pairs in OAM eigenstates. This is
prospective both for quantum communications and
optical-fiber-based metrology.

\acknowledgments The authors thank Juan P. Torres for his advice
and discussions as well as hospitality during the stay at ICFO.
Support by projects CZ.1.05/2.1.00/03.0058 and
CZ.1.07/2.3.00/20.0017 of M\v{S}MT \v{C}R and P205/12/0382 of GA
\v{C}R are acknowledged. D.J. and J.P. acknowledge support by
project PrF\_2013\_006 of IGA UP Olomouc. J.S. thanks the projects
CZ.1.07/2.3.00/30.0004 and CZ.1.07/2.3.00/20.0058 of M\v{S}MT
\v{C}R.

\pagebreak

\bibliography{javurek}

\end{document}